\documentclass[journal]{IEEEtran}
\usepackage[utf8]{inputenc}
\usepackage{graphicx}
\usepackage[numbers]{natbib}

\usepackage{xcolor}
\usepackage{subfigure}
\usepackage{url,amsmath,amssymb,amsfonts,longtable}
\usepackage{setspace}
\usepackage{url,amsmath,amssymb,amsfonts,longtable,ifpdf}
\usepackage{setspace}
\renewcommand{\baselinestretch}{1}
\usepackage{fancyhdr}
\usepackage{xspace}
\usepackage{todonotes}
\usepackage{caption}

\ifCLASSINFOpdf
\else
\fi
\begin{document}
%
\title{Fuzzy Commitments Offer Insufficient Protection to Biometric Templates Produced by Deep Learning}

%
%
%

\author{Danny~Keller,
        Margarita~Osadchy,
        and~Orr~Dunkelman
\thanks{All authors are with the department of Computer Science, University of Haifa, Israel; e-mails: dannkel@gmail.com,\{rita,orrd\}@cs.haifa.ac.il}}
\maketitle

\begin{abstract}
In this work, we study the protection that fuzzy commitments offer when they are applied to facial images, processed by the state of the art deep learning facial recognition systems. We show that while these systems are capable of producing great accuracy, they produce templates of too little entropy. As a result, we present a reconstruction attack that takes a {\em protected template}, and reconstructs a facial image. The reconstructed facial images greatly resemble the original ones. In the simplest attack scenario, more than 78\% of these reconstructed templates succeed in unlocking an account (when the system is configured to 0.1\% FAR). Even in the ``hardest'' settings (in which we take a reconstructed image from one system and use it in a {\em different} system, with different feature extraction process) the reconstructed image offers 50 to 120 times higher success rates than the system's FAR.
\end{abstract}

\begin{IEEEkeywords}
protection of biometric systems, entropy, deep face recognition, privacy
\end{IEEEkeywords}

%
\IEEEpeerreviewmaketitle

\section{Introduction}

Entrance control is based on three possible factors: what you know (passwords), what you have (devices), and what you are (biometrics)~\cite{Jain_intoToBiometrics}. While passwords have many well known problems (due to lack of entropy and the need to
support users who forget theirs passwords) and devices may be stolen or lost, biometrics is usually considered as a ``safe'' bet, as it is part of a person appearance or behaviour and is considered to be unique.


However, unlike passwords or stored keys which are consistent, the sampling of biometrics is always fuzzy: The capturing process contains two sources of ``noise'' --- the first has to do with the varying conditions (e.g., illumination, change of pose), which are unrelated to the identity. The second source of noise comes from the  changes in the biometric trait itself due to aging, injury, etc.  Decoupling undesirable variation from person's identity in an automatic way is very difficult and reduction in fuzziness  actually reduces the entropy of the biometric trait (even before the discussion on how much entropy there is in the biometric trait, to begin with). In other words, most biometric schemes and algorithms sacrifice entropy for improved robustness (i.e., accuracy of the authentication system).

The question of the amount of entropy a given biometric modality possess is an open research problem. Even the question of which entropy measure to use is complicated, e.g., Shannon's entropy is unsuitable (as most biometric traits are not naturally represented by binary strings). Moreover,  fuzziness in  biometric samples further complicates the entropy analysis, as the notion of ``similarity'' (i.e., two strings representing the same identity) is not based on equality but rather on a bounded distance (e.g., Hamming, Euclidean distance, or a set difference).

Some works tried to address the issue in specific contexts, e.g., the amount of entropy in specific representations (e.g.,~\cite{SimhadriSF19} in iriscodes) or an upper bound obtained due to a specific attack (e.g.,~\cite{BallardKR08,HidanoOT12,IgnatenkoW09,LimY16}).
As was discussed in~\cite{SutcuLM07}:
\begin{quote}
``The min-entropy of the original data cannot be easily determined, hence making it difﬁcult to conclude the key strength of the resulting system. Even the min-entropy of the data can be ﬁxed in someway, the entropy loss may be too large to be useful and it can be misleading.''
\end{quote}
These works, however, addressed older  representations that are characterized by poor robustness, which inevitably resulted in (a possibly large) entropy loss.

The introduction of deep learning gave hope that the problem can be solved --- having an efficient recognition with high entropy. The accuracy of deep learning-based systems is considered to be as good as humans, and in many cases, is claimed to outperform people. Indeed, deep learning, which is most popular in face domain, shows good robustness in large-scale tests~\cite{faceDL_survey}. One could hope that feature extraction based on deep networks could provide robustness (reduce fuzziness), while retaining the entropy due to identity. Unfortunately, the use of such representations does not resolve the problem, as we show in this work by a series of attacks that compromise the security and privacy of systems employing deep network representations.  

We start the paper by adapting the attacks of~\cite{HidanoOT12} against a fingerprint access system protected by fuzzy\footnote{The fuzzy commitment was later shown to be a special case of secure sketches~\cite{pinsketch}, and our results are easily extendable to secure sketches and even fuzzy extractors~\cite{pinsketch}.} commitment~\cite{FuzzyCommitment}  to the case of fuzzy commitment face based  systems that use recent deep learning technologies for template generation. Our findings show that one can indeed succeed in accessing accounts without having knowledge of the biometrics of the user.\footnote{In this paper we decided not to address the process by which the face is presented to the system. While many systems use liveliness detection to avoid presentation attacks, our settings are different, as we further discuss and explain in Section~\ref{sec:Model}.}

We then proceed by presenting our new attack that aims at extracting the raw biometric data from the protected template. After entering into the account, we take a step forward from the works of~\cite{HidanoOT12} and use information from the fuzzy commitment decoding process to obtain the actual binary template\footnote{As we later discuss, while systems based on deep learning rely on feature vectors, the fuzzy commitment scheme is applied to binary strings.} of the user. We then approximate the feature vector from which this binary template was obtained using a specially designed neural network. Next, we reconstruct a facial image from the approximated feature vector by applying NbNet~\cite{nbnet}. The attack thus show that fuzzy commitment applied to templates of facial images produced by recent DL, offers insufficient protection.

Finally, we show that the reconstructed image can be used to access the account of this person in other protected systems. When
both systems use the same enrolment image (as might happen when the user supplies a picture), the success of  accessing the account using the reconstructed image is more than 600 times higher than the false acceptance rate.\footnote{False acceptance rate is the rate in which the system accepts an imposter as the real user.} Moreover, we attack systems that rely on a different template extraction process and even a different enrollment image with success rate which is 60 times higher than the false acceptance rate. Moreover, we tested the reconstructed images against the Amazon Face Rekognition system, where 64.8\% of the reconstructed images passed the validation test (for 0.1\% FAR), when the reconstructed image was matched to the original image (or 44.6\% for matching the reconstructed image to a different image of the same user).
These attacks show that our reconstructed facial images resemble the original images in the context of authenticating the user (we discuss in Section~\ref{sec:Model} why this seems to be the only objective metric one can obtain). In other words, not only that the the user's privacy is compromised, also the security of other biometric systems is affected.

At this point, we remind the reader that the privacy of biometrics is a very serious matter. Legal considerations (such as the GDPR or the Illinois Biometrics Privacy act) and technical reasons (e.g., the hardness of changing the actual biometric) demand strong privacy guarantees for storing biometrics. Most notably, we wish to avoid the possibility of reconstructing the biometrics from the stored templates in the case of a data breach of the biomertic ``password'' file. Hence, our attack actually breaks privacy ``guarentees''  of modern deep learning facial recognition access systems protected by fuzzy commitment.

Furthermore, these attacks imply that using the current state of the art deep learning feature extraction together with fuzzy commitments does not satisfy {\em irreversibility} and {\em unlinkability}, two of the privacy properties required by many biometric systems. Irreversibility (see for example its definition in ISO 24745~\cite{ISO24745}) essentially suggests that an adversary that obtains a biometric template cannot extract the biometrics from it. Obviously, our attack invalidates this notion. The concept of unlinkability (again, see ISO 24745~\cite{ISO24745}) essentially suggests that an adversary that obtained biometric templates from two different systems cannot identify the shared users. Our attack allows reconstructing the image from the template stored in the first database, and checking whether it corresponds to a template in a second database. Hence, this second notion is currently not achievable by state of the art DL representations of facial images.

The rest of the paper is organized as follows: Section~\ref{sec:related_work} discusses previous related works. Section~\ref{sec:cryposystem} outlines the systems we attack, i.e., gives the general outline of a biometric authentcation system, which uses protected templates. Section~\ref{sec:Model} discusses the target for the adversary, her capabilities (the adversarial model), and the evaluation criteria for the success of the attacks.
In Section~\ref{sec:gusseing} we discuss the basic guessing attack that allows to unlock an account by ``guessing'' different faces (and offer an upper bound on the amount of entropy contained in binary templates without any protection). We run the guessing attack to recover the binary template, which we then use to approximate the true feature vector (as detailed in Section~\ref{sec:Reconstruction}), and subsequently  transform it into a facial image (using NbNet~\cite{nbnet}). The experimental verification of our attack is presented in Section~\ref{sec:exp}. Finally, Section~\ref{sec:summary} concludes the paper.

\section{Related Work}
\label{sec:related_work}

Our work develops and integrates the following components: adversarial guessing, attack on protected template, and reversing a biometric tempate.

\paragraph{Adversarial guessing} The concept of ``brute'' force attacks against systems is not a new one. Also in the context of biometrics, guessing attacks are well known. Previous works explored adversarial guessing as a method to empirically estimate the entropy offered by some representations. A guessing distance was defined in~\cite{BallardKR08} as a number of guesses required to obtain the most likely acceptable element according to the population feature distribution. However, approximating a feature distribution accurately is a challenging task. The work in~\cite{Nagar} focused on estimating the security against partial recovery of fingerprint features and defined a coverage effort curve as a number of guesses required for recovering a certain portion of the fingerprint minutiae.
A more generic entropy measure was proposed in~\cite{LimY16}. Similarly to guessing entropy~\cite{guessing_entropy}, they expressed the guessing effort as a function of a probability of usefully guessing at all possible trials and suggested two guessing attacks.
In~\cite{HidanoOT12} two empirical approaches were considered for evaluating quadratic Renyi entropy. First was a brute force attack using either bit-string templates of real people or code-words (assuming the knowledge of the system's parameters). The second approach assumed a knowledge of the user's commitment and  suggested narrowing down the search to templates that can be decoded using this commitment. In~\cite{NagarNJ12} a similar attack was analysed in terms of a minimum impostor decoding complexity.



Our work expands this line of research towards modern deep learning representations. In other words, the guessing attack is a way to upper bound the entropy offered by the feature vectors produced by the state of the art deep learning feature extraction tools.

\paragraph{Attack on a protected template} There is a large volume of work on attacking protected biometric systems (see the survey of~\cite{RathgebU11}): addressing various protection schemes, biometric traits, and configurations.
Previous work targetted the extraction of the (binary) template from the system (e.g.,~\cite{BRINGER15,Chang06,Rathgeb12}). In other words, once the template was reconstructed, the attack was declared successful.

In contrast, we look further and reconstruct the actual biometrics from the protected template. While this was done in the past for systems with no template protection~\cite{template_inverse_survey}, we are not aware of prior work that performed an end-to-end attack of reconstructing a biometric sample, e.g., facial image, from a protected template.

\paragraph{Reversing biometric templates} Methods that reverse biometric templates in unprotected systems  have been applied to obtain facial images~\cite{nbnet,Gomez-Barrero12,ZhmoginovS16}, fingerprints~\cite{Cappelli2007,Kim19,Ross}, iris~\cite{Galbally13,ICPR-2010-RathgebU,Venugopalan11} and other biometric traits (see the comprehensive survey of~\cite{template_inverse_survey} for details). Each of these works uses a different approach which depends on the biometric modality and the attacked system (i.e., the system which was used to obtain the templates from the biometric samples).

Our work relies on NbNet~\cite{nbnet}. NbNet trains a De-Convolutional Neural Network (D-CNN) 
to invert templates obtained by a CNN back to a facial image. NbNet comprises a cascade of multiple neighborly de-convolution blocks (NbBlocks). 
To address the issues of previous D-CNNs, such as noise and redundant channels, NbBlock learns a reduced amount of channels compared to the deconvolutional block, while the rest of the channels are learned using the already computed channels in the same layer. The resulting network was shown to perform well on the LFW and color FERET benchmarks.

NbNet can be applied to reverse templates obtained by any CNN feature extractor, as it uses it as a black box (without  the knowledge of the feature extraction algorithm). 
This method fits well in our attack, and we indeed make use of it to reconstruct facial images from reconstructed feature vectors.  

\section{Protected Biometric System Using Fuzzy Commitments}
\label{sec:cryposystem}

The system we consider is a biometric system for authentication (i.e., log-in). When the account is created, the user obtains a user name and submits her biometrics (in our case --- facial images) to the system. The biometrics are processed to produce a protected template, where the aim of the protection is to avoid an adversary from extracting the biometrics from the protected template in the case of a compromise. In subsequent uses, the user is presenting herself with the user name and her biometrics are sampled again (and compared with the protected template).

Generally speaking, the processes of enrollment and authentication (log-in) look in such protected biometric systems as follows:
\begin{itemize}
    \item Enrollment process:
    \begin{enumerate}
        \item The user presents raw biometric data using some sensor, and a feature extractor extracts a feature vector $v \in \mathbb{R}^d$ representing the obtained biometry.
        \item A binarization method is used to transform the feature vector $v$ into a bit string $b\in\{0,1\}^n$, where $n$ is some predefined bit string length.
        \item Fuzzy commitment is applied to $b$ using an $(n,k,2t+1)$ error correction code\footnote{The parameters of the code are: $n$ being the codeword length, $k$  the number of information bits and $t$ the maximal number of bit errors.} denoted as $\mathcal{C}$: a random code word $c\in\mathcal{C}$ is sampled, and a commitment $z$ is computed by $z = b\oplus c$.
        \item The client then uses some cryptographic hash function $h$ to calculate the hash $h(c)$ and sends it to the authorization server. The commitment $z$ is stored (usually in a user's token, but may be stored on the server side for usability).
    \end{enumerate}
    \item Verification process:
    \begin{enumerate}
        \item A user presents raw biometric data using some sensor, and after feature extraction obtains a feature vector $v' \in \mathbb{R}^d$.
        \item The same binarization method is used to obtain $b' \in \{0,1\}^n$ from $v'$.
        \item The client takes the stored commitment $z$ which belongs to the claimed identity, and obtains $c'$ by computing $c'=b'\oplus z$. Then $c'$ is passed through the ECC decoder, to receive $\hat{c}$. Note that if $c'\oplus c\leq t$, then $\hat{c}=c$ since $c\in\mathcal{C}$.
        \item The client calculates the hash $h(\hat{c})$ and sends it to the authorization server. The server compares $h(\hat{c})$ to the stored hash value belonging to the claimed identity, and determines whether the user is genuine (in case $h(\hat{c})=h(c)$) or an impostor (otherwise).
    \end{enumerate}
\end{itemize}

The three main components in the process are the feature extraction (discussed for deep learning systems in Section~\ref{sec:sub:FeatureExtraction}), binarization (discussed in Section~\ref{sec:sub:Binarization}), and fuzzy commitment~\cite{FuzzyCommitment} (discussed in Section~\ref{sec:sub:FuzzyCommitment}).

\subsection{Feature Extraction with Deep Networks}
\label{sec:sub:FeatureExtraction}

In this work we focus on facial images as the biometric modality. The state of the art in face recognition is based on deep convolutional neural
networks as they offer the best performing features for faces~\cite{faceDL_survey}.
We have used publicly available deep networks which ranked high in the BLUFR benchmark~\cite{blufr} for face recognition. The main one we used is the
Facenet feature extractor~\cite{facenet}. In addition, to validate that our results are significant (and some fluke of chance) and do represent some truth, we also used the LightCNN-29v2 network~\cite{LigthCNN} as well as the dlib library.\footnote{http://dlib.net/face\_recognition.py.html} The details about the networks are specified in Section~\ref{sec:exp}.

These systems accept a facial image and produce a feature vectors of real numbers. The feature vector spaces differ in their dimension and ranges. 
Yet, in most of the systems the decision whether two feature vectors are obtained from the same individual is based on the Euclidean distance or cosine similarity between the vectors. Systems using cosine similarity (as those in our experiments), deploy some threshold, where the similarity scores  above this threshold are considered ``same identity'', whereas the similarity scores below this threshold are considered ``different identities''. As a result, the threshold determines the False Acceptance Rate   (FAR)/False Rejection Rate (FRR).




\subsection{Binarization}
\label{sec:sub:Binarization}

Most biometric protection mechanisms were tailored for binary strings. Hence, the feature vector obtained from the feature extraction needs to be transformed into a binary string.\footnote{We note that binarization is also used in optimizing search in biometric databases.}
Various methods for binarization were studied with the main goal of keeping the performance of the original representations. While there are many possible binarization schemes, we opted to use a simple binarization based on random projections (used in over 400 constructions e.g.,~\cite{FengYJ10,JinGN06,RP1}).

The Random Projection (RP) binarization accepts a feature vector $v \in R^d$ and using a set of predetermined random projection matrix, $W\in R^{d \times d}$, produces a binary template $b\in \{0,1\}^d$ as follows:
\[
b=\frac{(sign(W^T v)+1)}{2}.
\]
While the matrix $W$ may be considered secret, in our settings we assume it is known to the adversary, as the adversary has access to the system's parameters as discussed in Section~\ref{sec:Model}. The matrix $W$ may be  user-specific or the same matrix for all users. For our attack, the results themselves (success rates) do not change if this is a user-specific matrix or the same matrix for all users, as we discuss in Section~\ref{sec:ftr_approximation}.\footnote{One thing that changes is that the inversion of binary template into a feature vector requires to train a new network for each user. This affects the time of the attack, but does not change the success rate of the attack.} Therefore, we assume that the same random projection matrix $W$ is used for all the users.


One aspect of binarization schemes is that they may degrade the accuracy of the system compared to feature vectors.
Namely, for the the same FAR, the true positive rates may drop. This in turn may make some of our attacks easier. However, as we show in the Section~\ref{sec:ex_binarization} measuring distance on the binarized strings obtained from random projections, maintains high accuracy rate (with a small loss). In other words, two images of the same individual are expected to have a small Hamming distance between the binary strings, whereas the Hamming distance between images of two different individuals is expected to be high (about half of the bits).


\subsection{Fuzzy Commitment}
\label{sec:sub:FuzzyCommitment}

The protection mechanism we study is the Fuzzy Commitment scheme~\cite{FuzzyCommitment} for template protection. This mechanism is widely used in many constructions related to biometrics, e.g.,~\cite{FengYJ10,Adamovic17,Kevenaar05,Kevenaar06,Yuan14} and Physically Unclonable Functions (PUFs), e.g.,~\cite{PUF1,PUF2}.

Fuzzy commitment~\cite{FuzzyCommitment} is a cryptographic primitive that allows precise and secure reconstruction of a noisy input. A noisy source is sampled once, and a binary string $b$ is generated. The aim is to generate a fuzzy commitment $z$, such that given $z$ and a new sample from the source $b'$ (assumed to be relatively close to $b$), the correct value of $b$ can be recovered. At the same time, the security requirement is that from $z$ it is impossible to learn any information\footnote{There are some relaxations of the ``impossible'' and ``any information''. For example, requiring that the amount of work necessary to learn information is infeasible. We disregard these relaxations as they do not affect our results.} about the value of  $b$.


The commitment procedure is based on error correction codes (ECC) and is done in the following way:
\begin{itemize}
    \item Given the input $b$, pick at random a codeword $c \in {\cal C}$, where $\cal C$ is an $(n,k,2t+1)$ error correction code.
    \item Set $z = b \oplus c$ (i.e., the shift from $b$ to $c$).
    \item Given $b'$ (the noisy input) and $z$, the value of $b$ can
    be computed by correcting $c' = z \oplus b$ to obtain $c$.
    \item From $c$ and $z$, the value of $b = c \oplus z$ can be easily recovered.
\end{itemize}
If the values of $b$ and $c$ are random, then $z$ does not reveal any information about $b$. Moreover, in some cases, it is possible to store the value of $h(c)$ to allow identifying whether the correct value of $c$ was recovered.

In the context of authentication systems, there is a way to know if the reconstruction was successful, as the user succeeds to login to the system only if $b'$ is sufficiently close to $b$. Hence, for the sake of this paper, we assume that $h(c)$ is also stored (possibly together with $z$), though our results hold even when $h(c)$ is not stored as long as the adversary has a method to identify whether the reconstructed  $b$ or $c$ are correct (e.g., by observing whether the login attempt succeeded).

\section{The Adversarial Model}
\label{sec:Model}

The adversary aims to recover facial images from the protected biometric template ($z$ and $h(c)$), i.e., to break the protection of the templates.
Before we turn into the actual attacks in Sections~\ref{sec:gusseing} and~\ref{sec:Reconstruction}, we discuss the adversarial settings in which our results hold. These settings are quite standard and common, and future works are expected to extend our results to weaker settings (e.g., from whitebox access to blackbox access).


\subsection{Adversarial Settings}

Throughout our attacks we assume the adversary has several capabilities:
\begin{itemize}
    \item {\it Whitebox access}: the adversary has knowledge of the biometric system that we attack. This assumption is a reasonable one in the context of commercial systems or academic ones. The adversary can always gain access to such systems in a legitimate manner (downloading the academic system or buying the commercial system).
    \item {\it Control of input/output}: The adversary can present to the system facial images and obtain the output of the system (either feature vectors in unprotected mode or protected biometric templates in the case of the protected ones).
    \item {\it Knowledge of the Feature Vectors}: The adversary knows what are the feature vectors produced by the system (denoted earlier by $v$) and the internal binarization process in use (i.e., the projection matrix $W$).

For academic systems, it is obvious that one can easily separate these two processes. For commercial ones, it may look as if a complete reverse engineering is needed. However, this is far from being the case, as the feature extraction procedure (especially in the deep learning world) is quite different than the binarization one, and they can be easily identified (even via a simple static analysis). In the case the adversary has obtained the source code (e.g., an open-source system, or as part of a commercial agreement with the vendor) or in cases where the system may produce both unprotected mode and protected mode (which allows the adversary to sample the feature vectors directly) this identification is much easier.

\item {\it Access to Templates}: The adversary has obtained the protected template from the system (or if the template is stored by the user, from the user).  This models an adversary that either broke into the system and exfilitrated its protected templates, or in scenarios when the system operator is a possible adversary. We note that as the helper data strings (i.e., $z$) are considered public in fuzzy commitments (and fuzzy extractors), we also assume that they are available to the adversary.
\end{itemize}

We assume that besides these, the adversary has no knowledge about who are the users of the system. This is motivated by the fact that we study the protection offered by the fuzzy commitment scheme for facial images.\footnote{%
An adversary who knows the identities of the users of the system can use social media sites~\cite{DBLP:journals/jpc/AcquistiGS14,DBLP:conf/acsac/PolakisLKMIKZ12} or real-life surveillance to obtain the users facial images. However, this attack vector (a) requires additional knowledge, and (b) is applicable to any protected biometrics system, whether it is a good protection or a bad one. Our approach allows testing whether the protection offered by the protection mechanism is sufficient.}

\subsection{Evaluating the Success of the Attack}
\label{sec:sub:Metric}


The success of the attack actually depends on the question --- {\em how good is the reconstructed image?} In other words, {\em how well it matches the real image of the user?} Given that biometric systems are fuzzy to begin with, one should not expect exact equality between the original template and the reconstructed one, but rather expect a similar ``enough'' result.

Obviously, ``enough'' is far from being a scientific measure. One can think of using a human judgement, i.e., looking at the images and determining whether they are of the same person or not. However, this approach does not offer consistent assessment.

We adopt the four scenarios defined in~\cite{template_inverse_survey} to the case of facial images. The better the reconstruction is, the better the success rate for these attacks (were a perfect reconstruction is expected to behave similarly to the FAR/FRR rates of the original system).

{\bf Same Image-Same Feature Extraction (SISFE):} This attack is essentially Type-I attack discussed in~\cite{nbnet} (and scenario~1 of~\cite{template_inverse_survey}). This attack tests whether the reconstructed image will be accepted by the targeted system or other systems that use the same template generation process (e.g., same product) and the same enrollment image (e.g., when the user supplies the image).


{\bf Different Image-Same Feature Extraction (DISFE):}
This attack is essentially Type-II attack discussed in~\cite{nbnet} (and scenario~2 of~\cite{template_inverse_survey}). The main difference between this attack and the previous one is that the reconstructed image is presented to the same (or similar) system, but for an account with a different enrollment image of the same user.

{\bf Same Image-Different Feature Extraction (SIDFE):} This attack
corresponds to scenario~3 of~\cite{template_inverse_survey}. It discusses the scenario in which the user supplied the same enrollment image to two systems. However, unlike the SISFE scenario, this time the systems use different feature extraction processes. The SIDFE settings thus study the transferability of the reconstruction between different systems (and different deep learning representations).


{\bf Different Image-Different Feature Extraction (FIDFE):}  
This attack, corresponding to scenario~4 of~\cite{template_inverse_survey}, is the most general one --- the systems use different DL networks for feature extraction, and the user was enrolled with two different facial images. We argue that if the reconstructed face succeed to unlock an account in a different system (with a different DNN) that stores a template obtained from a different facial image, then the reconstruction has indeed succeeded in capturing the ``representation'' of the user.

\section{Guessing Attack against Protected Deep Learning Systems}
\label{sec:gusseing}

Our first step is to unlock users' accounts with the most basic attack --- a guessing attack. We remind the reader that we assume that the adversary has the biometric ``password file'' and any helper data relevant to the users. Thus, security measures such as liveness detection are not relevant to the discussion and analysis.

As the system is an entry system, it should resist guessing attacks, i.e., the produced templates must have sufficient entropy (similarly to regular passwords). We note that template protection does not protect against a guessing attack and in some cases it may even increase its effectiveness (due to decrease in entropy resulting from the feature extraction and binarization processes).

Our guessing attack has two goals: First, it is the first step in the attack of Section~\ref{sec:binary_template_rec}, which recovers the binary template itself. Second, a guessing attack provides an upper bound to the security offered by the authentication system.

Our guessing attack is similar to that of~\cite{HidanoOT12}. The attack of~\cite{HidanoOT12} analysed an authentication system that used a naive representation (obtained from Gabor features over pre-aligned fingerprints~\cite{TuylsAKSBV05}). Our attack, on the other hand, targets representations obtain from modern deep learning networks (which are shown to be very robust~\cite{facenet,LigthCNN,VGG}) for the feature extraction module.


The attack starts by collecting many facial images, for example, by accessing a public source such as the internet. Each such facial image is then processed by the feature extraction and the binarization, and the binary template is stored in a database $DB$.
The attack proceed as follows. For the targeted account in the system, i.e., given $z,h(c)$ of some user, the adversary performs the following steps until the attack is completed.
\begin{enumerate}
    \item The adversary picks a string $b'$ from the databases $DB$ that she did not try yet.
    \item The adversary computes $c' = z \oplus b'$.
    \item The adversary applies the error correction procedure to $c'$ to obtain $\hat{c}$.
    \item The adversary checks whether $h(\hat{c}) = h(c)$. If so, she outputs $b'$ that unlocked the account. Otherwise, she goes back to Step~1 (unless the entire database was tested, in which case, she terminates).
\end{enumerate}

Our experiments in Section~\ref{sec:ex_guessing_attack} tested the guessing attack against a system with 300 users and a database $DB$ that was created from the faces of the other 5,449 identities of the LFW database. We found that the success of the attack depends on the choice of the feature extraction network. However, in all three tested feature types, for an FAR of\footnote{For a higher FAR of 1\%, the success rate is higher.}~0.1\%, our attack was successful in accessing over 75\% of the accounts, Moreover, in two of them it reached 96\% of the accounts (See Table~\ref{tab:1}).

Our simple guessing attack shows the practical outcome of the FAR. While the FAR of 0.1\% for a high true positive rate is considered as an excellent performance in various face recognition benchmarks (e.g. BLUFR~\cite{blufr}), it allows the guessing attack to unlock over 90\% of the accounts with an auxiliary set of 5,449 random faces unrelated to the users of the system. This points to a very poor security. Similar issue was discussed in~\cite{Daugman16} with respect to iriscode. Our experiments illustrate the same problem in \emph{modern deep learning facial system}, with the goal to highlight the common misconception about a secure FAR value. 

Moreover, we have observed a very interesting phenomenon in our experiments. While the average hit rate was similar to the FAR of the system (i.e., the chance that an image from the database $DB$ ``unlocks'' an account was similar to the false acceptance rate), the number of times a specific user's account was ``unlocked'' varied considerably. In other words, for some users it was significantly easier to access their accounts than for some others. This can be viewed as the fact that some faces are more ``common'' (like common passwords). For these users, the success rate of the attacks (see Section~\ref{sec:ex_guessing_attack}) is also higher.

\section{Reconstructing the Biometric Data from a Protected Template}
\label{sec:Reconstruction}

The guessing attack allows unlocking accounts. We now turn our attention to the bigger prize --- reconstructing a ``useful'' facial image from the protected template. Our attack contains three steps: 1) Obtaining the exact binary template (before the fuzzy commitment protection), 2) Approximating the original feature vector (prior to binarization) and 3) Reconstructing the facial image from the approximated feature vector. These steps are explained below.

\subsection{Obtaining the Binary Template}
\label{sec:binary_template_rec}

We denote the genuine enrolled feature vector of the compromised user as $v$, its corresponding binary vector as $b$ and its fuzzy commitment as $z$. Our aim is to extract $b$ using the binary template $b'$ from $DB$ that unlocked the account.

Using $b'$ and $z$, we can easily obtain the genuine $b$: First, we compute $c'=b'\oplus z$. Then, we decode $c'$ using the ECC to obtain $\hat{c}$. Finally, we compute $\hat{b}=\hat{c}\oplus z$.

Given that the account was unlocked then $\hat{b}=b$, and we are done. This follows immediately from the verification process described in Section~\ref{sec:cryposystem}: access to the account suggests that the computed $\hat{c}$ is correct (as $h(\hat{c})=h(c)$). This can happen only\footnote{Of course, there is a small chance of a hash function collision. However, if $h(\cdot)$ is a good cryptographic hash function, the probability of this event is negligible.} if $c' = b' \oplus z$ was corrected to $c=z \oplus b$. As the adversary knows both $\hat{c}$ and $c'$ she can deduce the bits in $b'$ that need correction to obtain $b$ (as $b = \hat{c} \oplus c' \oplus b'$).

\subsection{Approximating the original feature vector}
\label{sec:ftr_approximation}

The next step is to ``reverse'' the binarization, i.e., to find a real feature vector $\hat{v}$ that yields the binary vector $b$, if undergone the system's binarization process. If the binarization process is indeed good, then $\hat{v}$ is expected to be close to the user's supplied $v$ (a fact which we use in Section~\ref{sec:sub:CallingNbNet}).

We note that this step is the only one in our attack that depends on the choice of the binarization matrix $W$. Hence, if the system uses a different $W$ per user, then this step need to be run for each user separately. This does not impact the success rate of the attack (but does require training the networks again, which takes more time).

Our task in this step is to invert the binarziation process. Let $f$ denote the binarization process: $f(v)=b$. We seek to find $\hat{v}= argmin_{x\in V}  \;\;dist_H(f(x),b)$, where  $dist_H$ is the Hamming distance and $V$ is the feature space.
There are several problems with this optimization formulation. First, both $f$ and $dist_H$ are not differentiable; second, this formulation usually has many solutions, and most of them may be far from the original feature vector $v$.

Let $g$ denote the reverse function that maps the binary template back to its original feature vector. We address both problems by defining $f$ and $g$ as neural networks and forcing cyclic consistency between them as detailed below (cyclic consistency loss has been widely used in image transfer applications~\cite{ZhuPIE17}.)


\subsubsection{Loss Functions and Training of the Networks}\label{sec:losses}

Let $F:v \rightarrow b$ denote the network that approximates the binarization process and let $G: b \rightarrow v$ denote the network that approximates the reverse mapping from the binary template  to the real feature vector (the one that we seek in this step of the attack).
We train the networks $F$ and $G$ on (a training) set of pairs  $(v_i,b_i)$ obtained from applying the feature extraction and binarization process of the target system on a public set of facial images.
The networks are trained simultaneously, namely, in each iteration we run the feed-forward pass in both netwokrs to obtain the estimates and then update the parameters of the networks. Each network is trained with a loss function that combines its prediction loss and two consistency losses
\begin{equation} \label{eq:1}
    L = L_{pred} + \lambda \cdot (L_{cyc-ftr} + L_{cyc-bin})
\end{equation}
where $L_{pred}=L_F$ for training the binarization network $F$ and $L_{pred}=L_G$ for training the reverse network $G$;
 \(\lambda\) is a parameter for controlling the weight of the cycle consitency losses in the learning process.

\paragraph{$L_F$ - prediction loss for network $F$} We want to minimize the Hamming distance between the prediction $F(v)$ and the ground truth $b$ which was produced by the binarization process on $v$. Since Hamming distance is not differentiable, we approximate it with the binary cross entropy (BCE) loss, which treats each bit independently.
\paragraph{$L_G$ - prediction loss for network $G$} We define $L_G$ as the minimum square error (MSE) loss between the prediction $G(b)$ and the original feature vector $v$ (which served as an input to  the binarization process that produced $b$).
\paragraph{$L_{cyc-ftr}$ - feature cyclic consistency loss}  To reduce the space of possible mapping functions, the learned mapping should be cycle-consistent:  for each feature vector $v$, the transformation cycle from the Euclidean space to the Hamming space and back should be able to bring the mapping $G(F(v))$ back to the original feature vector $v$, namely $v\rightarrow F(v)\rightarrow G(F(v)) \approx v$. To enforce this consistency we define the feature consistency loss as MSE loss between $G(F(v))$ and $v$.
\paragraph{$L_{cyc-bin}$ - binary template cyclic consistency loss} Similarly to $L_{cyc-ftr}$, we define the binary template consistency loss to ensure that $b\rightarrow G(b)\rightarrow F(G(b)) \approx b$. We define $L_{cyc-bin}$ as BCE loss between $F(G(b))$ and $b$.

\subsubsection{Network Architecture}
The feature vectors and the binary templates in our experiments are both of dimension 128. Thus we use the same arhitecture for both networks $F$ and $G$. The architecture comprises input and output layers of size 128 and two fully connected hidden layers of size 256, all with sigmoid activation. Due to sigmoid function, which operates as a gate, the output of the binarization network is close to binary. Thus we pass $F(v)$ directly to network $G$ without binarizing it.

\subsection{Reconstructing the Face}
\label{sec:sub:CallingNbNet}
Once we obtain the approximated $\hat{v}$ we can reconstruct the facial images from it. For that purpose we use the pre-trained NbNet~\cite{nbnet} to reconstruct a facial image from $\hat{v}=G(b)$.

As we show in Section~\ref{sec:exp}, in many cases, the reconstructed image is sufficiently close to that of the user to allow unlocking his account (even at 0.1\% FAR) and it is even visually similar to it (see Figure~\ref{tab:iill:1}). This suggests that with very high probability, the reconstruction is successful, and the reconstructed image can be used to infer private information about the user, e.g., gender, age, or ethnicity. Finally, as we discuss in Section~\ref{sec:summary}, this suggests that the fuzzy commitments applied to deep-learning representation does not offer irreversibility or unlinkability (as one can reconstruct the image of the user from one protected system, and try to use it to unlock the account in the second system).

Finally, we note that NbNet was trained for FaceNet feature vectors, but it is not the only work on recovering faces from the feature vectors (see for example,~\cite{DBLP:conf/bmvc/MignonJ13,DBLP:journals/pami/MohantySK07}). Hence, the use of the NbNet is merely a matter of convenience, and can be replaced with any reconstruction tool. 

\section{Experiments and Results}
\label{sec:exp}

We  report the implementation details of the target system in Section~\ref{sec:ex_target_sys}  and the details of the neural networks training in Section~\ref{sec:ex_NN_training}. Following the works of~\cite{emplate_inverse_survey}, Section~\ref{sec:ex_attack_analysis} provides the evaluation of our attack under the four different metrics defined in Section~\ref{sec:sub:Metric}. We also performed the SIDFE and DIDFE tests (the hardest ones) using Amazon's Face Rekognition, and report the results in Section~\ref{sec:sub:Amazon}. 



\subsection{Implementation Details}
\subsubsection{Target System}\label{sec:ex_target_sys}
Our target system is an authentication system that operates on facial images. The input image is converted into a real feature vector of size 128 using Facenet deep network~\cite{facenet}.\footnote{We used the implementation from \url{https://cmusatyalab.github.io/openface/}.} It is then transformed to a binary vector of size 128 as described in Section~\ref{sec:sub:Binarization} using the same projection matrix $W$ for all users.
The system applies fuzzy commitment to the binary vector and saves the commitment and the hash of the codeword in the system.\footnote{We used the implementation of Pinsketch~\cite{pinsketch} provided in
 \url{https://www.cs.bu.edu/~reyzin/code/fuzzy.html}. Since Pinsketch is  designed for the set difference, we converted the binary vector $b$ to a set by listing the coordinates of the bits which are which are set to~1.}
 In the verification phase, the input image follows the same feature extraction and binarization processes. Then the system tries to decode the codeword with the corresponding commitment and compares its hash to the one stored in the system. The user is granted  access if they are identical and denied access otherwise.  We use the BLUFR benchmark \cite{blufr} to determine the system's thresholds for FARs of 1.0\% and 0.1\%.\footnote{BLUFR is a benchmark for face verification built for the LFW dataset focusing on low FPRs. It consists of 10 sessions, each containing about 156,915 genuine matching scores and 46,960,863 impostor matching scores on average for performance evaluation.}

\paragraph{Users}  LFW deep-funneled \cite{LFW-df} data set includes 13,233 face images of 5,749 identities and it is one of the most popular benchmarks for face verification.  Out of the 1,680 identities which have two or more images, we picked 300 identifies as users of the system. The remaining 5,449 were used as an auxiliary set as detailed in Section~\ref{sec:ex_attack_analysis}.

\subsubsection{Training the Feature Vector Recovery Neural Network}
\label{sec:ex_NN_training}
To train the network for approximating a feature vector of the target user from a binary template, we used a data set comprising 35,000 images from the VGG-Face dataset \cite{VGG}.
We obtained feature vectors and binary templates for these images using the feature extraction process and binarization of the target system (Section~\ref{sec:ex_target_sys}). The feature vectors served as inputs for training the binarization network $F$ and their corresponding binary templates served as targets. The input and targets were reversed for training the feature recovery network $G$. Both networks where trained simultaneously using a stochastic gradient descent (SGD). We set the \(\lambda\) parameter in the loss function (Eq.~\ref{eq:1}) to $0.85$. We set the learning rate to $0.9$ and the batch size to $50$. The trained network was tested on almost 6,000 unseen images. The similarity between the recovered and real feature vectors were above the threshold corresponding to 0.1\% FAR for all tested images. 


\subsection{Analysis of the Attack}~\label{sec:ex_attack_analysis}
Our attack uses template guessing to obtain the binary feature vector. Then, this binary feature vector is used to find an approximate feature vector, which in turn is used for image reconstruction. The reconstruction part can be performed for the accounts that were successfully accessed by the guessing attack. We perform the full attack on the target system comprising 300 users (as detailed above).

\subsubsection{Guessing Attack}\label{sec:ex_guessing_attack}
Our guessing attack, detailed in Section~\ref{sec:gusseing}, uses an external set of facial images (which could be any public set.) In our experiments we used an auxiliary set comprising facial images of 5,449 identities from LFW (one image per person)  that were not used as users of the target system. 289 out of 300 accounts were authenticated as genuine users using the proposed guessing attack.

To evaluate the robustness of the guessing attack to the choice of the feature extraction method, we ran the attack in the same settings with two additional modern face networks: LightCNN29v2~\cite{LigthCNN} and Dlib~\cite{dlib-resnet}.\footnote{The feature vector length in the original extraction network (FaceNet) and in Dlib is 128, and in LightCNN is 256. We used the same lengths for the binary vectors.} The results of the original feature extraction and the two additional networks are summarized in Table~\ref{tab:1}. The attack was performed in both unprotected and protected settings, while in both cases a threshold corresponding to 0.1\% FAR was used (a cosine similarity threshold for the unprotected setting domain, and a hamming distance threshold for the protected setting domain.)

\begin{table}[h!]
    \centering
    \begin{tabular}{||l c c ||}
        \hline
        Feature extraction  & Protected templates & Feature vectors \\[0.5ex]
                \hline\hline
        FaceNet~\cite{facenet}  & 96.3 \% & 86.0 \%\\
        \hline
        Dlib~\cite{dlib-resnet} & 75.0 \% & 87.3 \% \\
        \hline
        LightCNN~\cite{LigthCNN}  & 96.0 \%& 85.3 \%\\
        \hline
    \end{tabular}
    \caption{Success rate of Guessing Attack (for an FAR of 0.1\%)}
    \label{tab:1}
\end{table}


Next, we analyzed the success of the guessing attack in terms of probability of hitting a user. For this test we chose at random 5\% of the LFW subjects as the target set and the rest as the auxiliary set.  Figure~\ref{fig:2} (orange bars) shows the histograms of the success probability for the three tested feature extraction networks. For each target, the success probability is namely the ratio of images from the auxiliary set that were able to access it. While the FAR is 0.001, these results demonstrate that the guessing attack is much more successful than FAR for a significant amount of the users (Y axis).

\subsubsection{Assessing the Impact of Binarization}\label{sec:ex_binarization}
\begin{figure}[h!]
    \centering
    \subfigure[]{\includegraphics[width=0.5\textwidth]{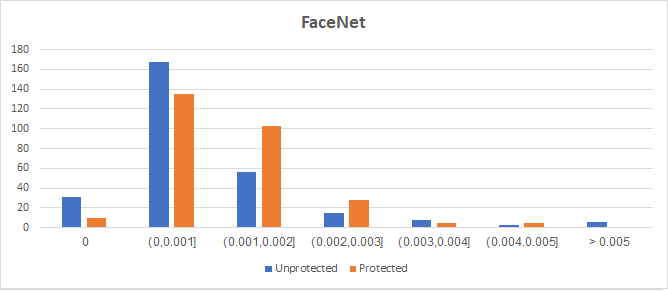}} 
    \subfigure[]{\includegraphics[width=0.5\textwidth]{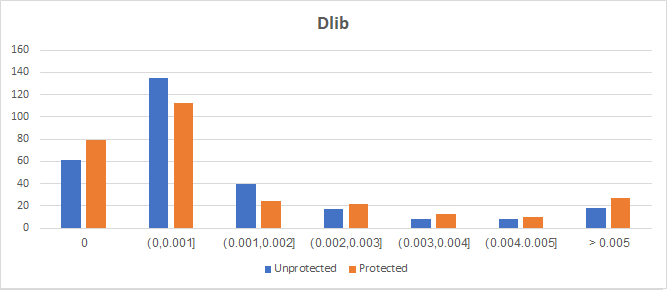}}
    \subfigure[]{\includegraphics[width=0.5\textwidth]{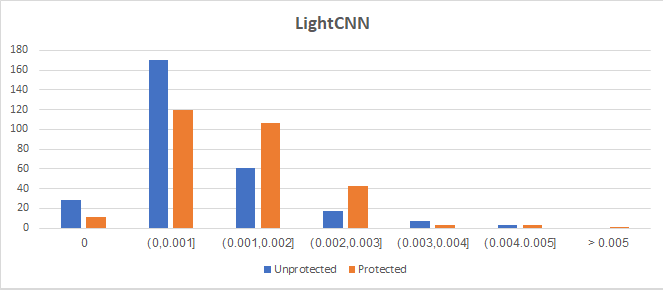}}
    \caption{Probability of hit distributions before (in blue) and after (in orange) binarization: (a) FaceNet (b) Dlib (c) LightCNN.}
    \label{fig:2}
\end{figure}

As described in Section \ref{sec:sub:Binarization}, we use random projections to transform the face templates from the real-valued feature space to  binary strings. One can argue, that if our  binarization methods reduces the entropy significantly, the proposed attack would not be effective for a better binarization method. To check this we compared the results of the guessing attack on binary strings (obtained using random projection binarization)  and feature vectors (no binarization -- no entropy loss).
As shown in Figure \ref{fig:2}, the distribution of hit probabilities in binarized (orange) and the real-valued (blue) domains are similar.

Another way to assess the efficiency of the attack is by assuming ``perfect binarization'' that does not affect the accuracy of the original feature vector. To imitate a perfect binarization we defined the success of the attack if the cosine similarity between the normalized feature vector  of the candidate face and normalized feature vector extracted from the user's face was over the system's threshold (corresponding to FAR of 0.1\%). We observe in Table~\ref{tab:1} that even for this ideal case (in terms of entropy loss), the success of the guessing attack is still very high, over 85\% in all tested face networks.

\subsubsection{Reconstruction Attack}
\label{sec:sub:ReconstructionResults}

The reconstruction attack uses the binary template obtained from the fuzzy commitment of the guessed account (as described in Section~\ref{sec:Reconstruction}) as an input to the network $G$, which outputs the approximation of the user's feature vector. We pass this feature vector to NbNet~\cite{nbnet}  to reconstruct a facial image of the user. The reconstructed image is the final goal of the proposed attack.

NbNet~\cite{nbnet} has shown high success in reconstructing facial images from unprotected feature vectors. Since our approximation of the  feature vector is close but not equal to the original feature vector,  we cannot guarantee that the image reconstructed by NbNet from the approximated feature vector matches the success of the image recovered from the original feature vector. Next, we evaluate the usefulness of the reconstructed image in a number of settings.


Following the metrics proposed in~\cite{template_inverse_survey} and detailed in Section~\ref{sec:sub:Metric}, we test the success rate for two thresholds yielding  1.0\% and  0.1\% FAR.  We also report the results of the same test carried out using the NbNet reconstruction of the original feature vector (instead of the recovered one). Since we use this technique to reconstruct images, our success rate is naturally bounded by the level of success of the original reconstruction. Additionally, we report the performance of the system itself. While in SISFE and SIDFE this is not necessary (obviously all the feature vectors are identical to themselves), in DISFE and DIDFE, we use the system performance with regards to its level of separation between genuine and impostor images, as an anchor to which our results can be compared. All results are reported for the feature space and for the binary templates (assuming the system is protected). 


\begin{figure*}[tb]
\centering
      \includegraphics[width=1.22 cm]{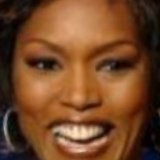} 
      \includegraphics[width=1.22 cm]{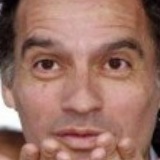} 
      \includegraphics[width=1.22 cm]{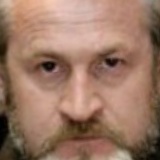} 
      \includegraphics[width=1.22 cm]{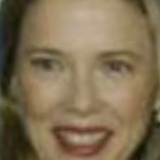} 
      \includegraphics[width=1.22 cm]{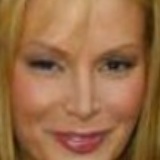} 
      \includegraphics[width=1.22 cm]{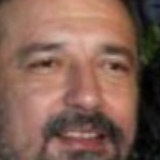} 
      \includegraphics[width=1.22 cm]{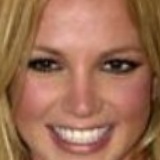} 
      \includegraphics[width=1.22 cm]{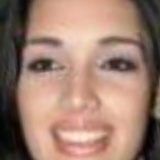} 
      \includegraphics[width=1.22 cm]{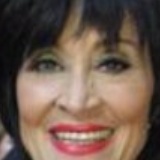} 
      \\
    \includegraphics[width=1.22 cm]{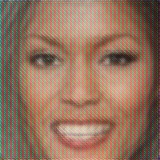}
      \includegraphics[width=1.22 cm]{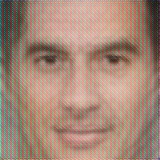}
      \includegraphics[width=1.22 cm]{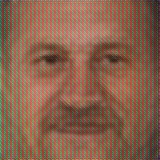}
      \includegraphics[width=1.22 cm]{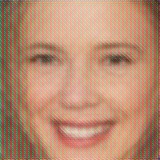}
      \includegraphics[width=1.22 cm]{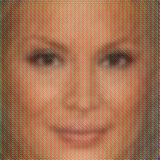}
      \includegraphics[width=1.22 cm]{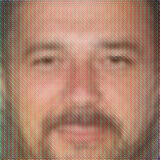}
      \includegraphics[width=1.22 cm]{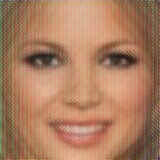}
      \includegraphics[width=1.22 cm]{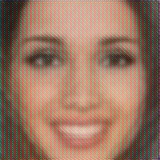}
       \includegraphics[width=1.22 cm]{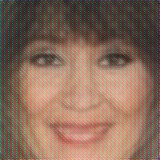}
      \\
       \includegraphics[width=1.22 cm]{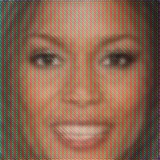}
      \includegraphics[width=1.22 cm]{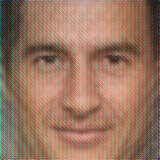}
      \includegraphics[width=1.22 cm]{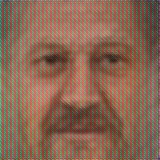}
      \includegraphics[width=1.22 cm]{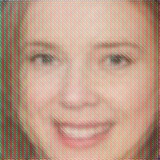}
      \includegraphics[width=1.22 cm]{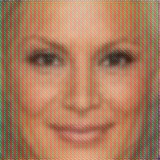}
      \includegraphics[width=1.22 cm]{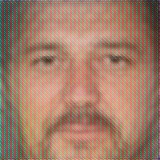}
      \includegraphics[width=1.22 cm]{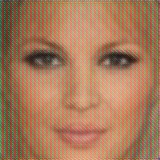}
      \includegraphics[width=1.22 cm]{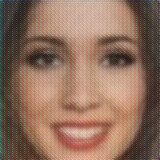}
      \includegraphics[width=1.22 cm]{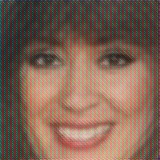}
       \\
       \setlength{\tabcolsep}{2pt}

       \begin{tabular}{p{1.22cm}p{1.22cm}p{1.22cm}p{1.22cm}p{1.22cm}p{1.22cm}p{1.22cm}p{1.22cm}p{1.22cm}}

       {\small {\bf SISFE}}& {\small {\bf SISFE}} & {\small {\bf SISFE}} &{\small {\bf SISFE}} & {\small {\bf SISFE}} & {\small {\bf SISFE}} &  {\small {\bf SISFE}} & {\small {\bf SISFE}} & {\small {\bf SISFE}}\\
       {\small {\bf DISFE}} & {\small {\bf DISFE}} & {\small {\bf DISFE}} &{\small {\bf DISFE}} & {\small {\bf DISFE}} & {\small {\bf DISFE}} & {\small {\bf DISFE}} & {\small {\bf DISFE}} \\
       {\small {\bf SIDFE}} & {\small {\bf SIDFE}} & {\small {\bf SIDFE}} &{\small {\bf SIDFE}} & {\small {\bf SIDFE}}  \\
       {\small {\bf DIDFE}}  \\
       \end{tabular}

    \caption{Examples of reconstructions: top row --- original image, middle row --- reconstructed from the template, bottom row --- reconstructed from the protected template. All reconstructed images passed all four tests for an FAR of~1\%. Under each reconstructed image we list the tests it has passed for an FAR of~0.1\%}
    \label{tab:iill:1}
\end{figure*}


\paragraph{SISFE Results}
In SISFE evaluation, the same image is used in the test, as well as the same feature extraction process.  
The results of the SISFE evaluation are given in Table~\ref{tab:SISFE}. The results show that usually the images we recover from the protected templates are very similar to the enrollment images. Some examples of the reconstructed images obtained by our attack are shown in Figure~\ref{tab:iill:1} in comparison to the target real image and the reconstruction obtained from the real template.

As similarity is a subjective notion, we use the SISFE metric that suggests that the reconstructed images are sufficiently close in the feature vector space to the enrollment image. Hence, they can be used to obtain access for different systems that use the same feature extraction process and the same enrollment image as the target system. The reported success rates regard the reconstruction phase, while the results in parenthesis regard the success rate of the entire attack, including the guessing attack stage. It also could be seen that the reconstruction attack performs better than the original reconstruction in the binary space. This could be explained by the fact that the initial point of the attack was the binary vector, thus by transforming the feature vector extracted from the reconstructed image, to the binary space, we are essentially getting closer to the original vector. With the original reconstruction, it is the other way around --- as the initial point of the reconstruction was the genuine feature vector,  moving to the binary space get us further away from the original vector.

\begin{table*}[tb]
    \centering
    \begin{tabular}{||l c c c ||}
        \hline
       Image Type& Binarization & 1.0\%  FAR& 0.1\% FAR\\ [0.5ex]
           \hline\hline
        Original&N& 100\% & 100\% \\
        \hline
        Original&Y& 100\% & 100\% \\
        \hline
        Reconstructed from unprotected template& N& 98.96\% (95.36\%)  & 95.16\% (95.36\%)  \\
        \hline
        Reconstructed from protected template& N & 96.19\% (92.69\%)  & 83.39\% (80.36\%) \\
        \hline
        Reconstructed from unprotected template & Y & 94.81\% (91.37\%) & 81.31\%(78.34\%) \\
        \hline
        Reconstructed from protected template& Y & 96.19\% (92.69\%)&  87.89\% (84.70\%) \\
        \hline
    \end{tabular}
    \caption{SISFE evaluation results of the reconstruction attack. Success rates of the full attack (including the guessing attack) are reported in parenthesis.}
    \label{tab:SISFE}
\end{table*}

\paragraph{DISFE Results}
In DISFE evaluation, we examine the similarity between the reconstructed image and a different image of the same user, but we assume the same feature extraction process. 
For this test we used an additional image of each user (the subjects selected for the users set had at least two images in LFW).
Table \ref{tab:DISFE} shows the results of DISFE evaluation. Changing the enrolment image renders the problem more complex, as in this setting the attack must generalize to different appearances of the subject and not only to a single image. Note, that the variation of the appearance in images of the same subject in LFW is very large, including changes in pose, illumination, age, hair style, etc. 
In addition, the accuracy of the NbNet (for unprotected templates) drops significantly. Still, our reconstruction attack reaches a success rate of 66\% for 1.0\% FAR threshold and more than 35\% for the 0.1\% FAR threshold.

\begin{table*}[tb]
    \centering
    \begin{tabular}{||l c c c ||}
        \hline
       Image Type& Binarization & 1.0\%  FAR& 0.1\% FAR\\ [0.5ex]
           \hline\hline
        Original&N&  99.11\% & 97.34\% \\
        \hline
        Original&Y&  95.44\% & 85.94\%  \\
        \hline
        Reconstructed from unprotected template& N& 87.89\% (84.70\%)  & 65.05\%  (62.69\%) \\
        \hline
        Reconstructed from protected template& N & 76.12\% (73.36\%)& 41.18\% (39.68\%)\\
        \hline
        Reconstructed from unprotected template& Y & 65.74\% (63.35\%)& 46.02\% (44.35\%) \\
        \hline
        Reconstructed from  protected template& Y & 66.44\% (64.03\%) & 36.68\%  (35.35\%)\\
        \hline
    \end{tabular}
 \caption{DISFE evaluation results of the reconstruction attack. Success rates of the full attack (including the guessing attack) are reported in parenthesis.}
    \label{tab:DISFE}
\end{table*}



\paragraph{SIDFE Results}
In the SIDFE evaluation, we assume that the feature extraction process in the new system is different from the target system, but the user used the same image to enroll to both systems. We evaluated two feature extraction networks, specifically, LightCNN and Dlib. We do not assume any knowledge of the binarization procedure in the new system.

The results of the authentication in the feature space  correspond to the case in which the binarization in the new system does not affect the accuracy and thus can serve as an upper bound. We report the results with the binarization for completeness and keep the same binarization method as in the target system for simplicity (any other methods could be used instead).

The results in Table~\ref{tab:SIDFE} show that the reconstructed images are sufficiently close to the original ones.
Our attack reconstructs images of users that are very useful (in the range of 40\%--56\% with no binarization  and in the range of 13\%--22\% with  binarization).

\begin{table*}[tb]
    \centering
    \begin{tabular}{||l c c c c c||}
        \hline
         Image Type& Binarization &  1.0\% FAR  & 0.1\% FAR  & 1.0\% FAR & 0.1\% FAR\\
         & & LightCNN &LightCNN&dlib&dlib\\ [0.5ex]
        \hline\hline
        Original&N& 100\% & 100\% & 100\% & 100\% \\
        \hline
        Original&Y& 100\% & 100\% & 100\% & 100\% \\
        \hline
        Reconstructed from & N& 91.00\%  & 72.66\% &91.35\% & 78.89\%\\
        unprotected template& &(87.70\%)&(70.03\%)&(88.03\%) &(76.03\%)\\
        \hline
        Reconstructed from & N & 73.01\% & 39.79\% &78.55\%& 55.71\%\\
        protected template & & (70.36\%)&(38.35\%))&(75.70\%) &(53.68\%) \\
        \hline
        Reconstructed from & Y &74.39\% & 43.60\%  & 50.17\%&22.84\% \\
        unprotected template& & (71.69\%) & (42.04\%)&(48.35\%) &(22.01\%) \\
        \hline
         Reconstructed from & Y & 47.40\%  & 21.80\% & 37.02 &13.15\%\\
        protected template & &(45.68\%) & (21.01\%) &(35.68\%)&(12.69\%)\\
        \hline
    \end{tabular}
    \caption{SIDFE evaluation results of the reconstruction attack. Success rates of the full attack (including the guessing attack) are reported in parenthesis.}
    \label{tab:SIDFE}
\end{table*}

\paragraph{DIDFE Results}
In DIDFE evaluation, both the feature extraction process and the enrolment image are different from that of the target system. The feature extraction and binarization in the new system are unknown during the reconstruction attack. We use the same template generation process as in SIDFE evaluation setting and we use an additional image of the user for enrollement in the new system (as in DISFE evaluation). The results are shown in Table~\ref{tab:DIDFE}. In this most challenging setting our attack is capable of reconstructing images that are useful for accessing the accounts of over 20\% of the users with no binarization and 5\%--12\% with binarization (at 0.1\% FAR).

\begin{table*}[tb]
    \centering
    \begin{tabular}{||l c c c c c||}
        \hline
        Image Type& Binarization &  1.0\% FAR  & 0.1\% FAR  & 1.0\% FAR & 0.1\% FAR\\
         & & LightCNN &LightCNN&dlib&dlib\\ [0.5ex]
        \hline\hline
        Original & N&98.45\% & 95.66\% & 98.29\% & 92.23\% \\
        \hline
        Original& Y& 94.64\% & 84.08\% & 49.82\% & 20.21\% \\
        \hline
         Reconstructed from & N&  65.05\%  & 38.75\% &77.16\% & 49.83\%\\
        unprotected template& &(62.69\%)&(37.34\%)&(74.37\%) &(48.02\%)\\
        \hline
         Reconstructed from & N & 46.37\% &  20.76\% &64.71\%& 26.30\%\\
        protected template & & (44.69\%)&(20.01\%))&(62.36)\% &(25.35\%) \\
        \hline
        Reconstructed from & Y & 50.52\% & 20.42\%  & 29.76\%&11.07\% \\
        unprotected template& & (48.69\%) & (19.68\%)&(28.68\%) &(10.66\%) \\
        \hline
        Reconstructed from & Y & 33.22\%  & 12.46\% & 23.18 &5.19\%\\
        protected template & &(32.01\%) & (12.00\%) &(22.36\%)&(5.00\%)\\
        \hline
    \end{tabular}
    \caption{DIDFE evaluation results of the reconstruction attack. Success rates of the full attack (including the guessing attack) are reported in parenthesis.}
    \label{tab:DIDFE}
\end{table*}

\subsection{Using Reconstructed Images in a Commercial System}
\label{sec:sub:Amazon}

In order to get a sense of the consistency and quality of the reconstructed images obtained using our attack, we performed the SIDFE and DIDFE evaluations using the Amazon Face  Rekognition\footnote{https://docs.aws.amazon.com/rekognition/latest/dg/faces-comparefaces.html} commercial system. We note that we have no knowledge about any part of this framework.

We have generated an artificial authentication system based on the Rekognition framework. First, we assumed that the system enrolls
users with a single facial image. For subsequent authentication attempts, the Amazon Rekognition was invoked with the original enrollment image and the new sampled one of the subject. To decide whether the verification succeeded, we had to decide on similarity score that suggests ``acceptance''. We ran our tests on LFW dataset, but due to limited resources, we were not able to compute threshold on the Amazon's verification score using the BLUFR benchmark. 
Instead, we chose at random  1000  genuine and 1000 impostor pairs from the BLUFR first test~\cite{blufr} to found the threshold corresponding to 0.1\% FAR. For that threshold, the system yielded 99.1\% true positive rate.

After setting the threshold, we applied the attack of Section~\ref{sec:Reconstruction} in the two settings (SIDFE and DIDFE) for 1000 individuals chosen randomly from LFW. For the SIDFE evaluation of the 1000 reconstructed images, we obtained a 64.8\% success rate. The DIDFE evaluation produced 44.64\% success rate for the FAR rate of 0.1\%. These results show that our attack is capable of consistently reconstructing facial images from protected templates. These reconstructed images succeed in tricking black-box state-of-the-art commercial systems with high probability.


\begin{figure}[h!]

\end{figure}


\section{Conclusions}
\label{sec:summary}

In this paper we studied for the first time whether fuzzy commitment can protect deep-learning presentations for facial images. The clear answer is that current state of the art deep learning representations do not carry enough entropy. Thus, they are susceptible to guessing attacks that allow recovering a related facial image from the protected template. Our results were tested in different working environments, showing success rates which are significantly higher (by orders of magnitude) over the FAR of the system.

This suggests that fuzzy commitment is insufficient to offer either irreversibility or unlinkability for facial images. Besides the immediate impact on the claimed security of systems that use deep learning features for facial images, this may also impact cancelable biometrics solutions which are based on fuzzy commitments --- once the approximated facial image is reconstructed, the adversary can enroll it on his own (similarly to the SIDFE attack scenario).

Future works may decide to extend the applicability of our results. For example, we currently rely on white-box access to the feature extraction process, but it seems mostly a matter of training the correct neural network to transform the attack to black-box settings. Another line of research might be applying these results to other modalities, e.g., finger prints, as deep learning-based feature extraction processes are becoming prevalent (see for example~\cite{DBLP:journals/corr/abs-1904-01099}).


%



\section*{Acknowledgements}

The authors would like to thank one of the authors of \cite{nbnet}, Guangcan Mai, for his generous assistance in providing clarifications and material regarding the work published in~\cite{nbnet}, during our research. We are also grateful to the support offered by the Identity and Biometric Applications Unit in Israel's National Cyber Directorate.

This research was supported in part by the Israeli Ministry of Science and Technology through project 3-14659 and by the Center for Cyber, Law and Policy at the University of Haifa with the Israeli Cyber Directorate.

\bibliographystyle{IEEEtran}
\bibliography{ref}

\end{document}